\documentstyle[prl,aps,twocolum,psfig]{revtex}

\begin{document}
\hoffset=-0.5in
\textwidth=7in

\preprint{LBNL-41726, DOE/ER/40561-5-INT98}
\twocolumn[\hsize\textwidth\columnwidth\hsize\csname @twocolumnfalse\endcsname

%

\title{Effects of Jet Quenching on High $p_T$ Hadron Spectra in 
High-energy Nuclear Collisions}
\author{Xin-Nian Wang}
\address{
Nuclear Science Division, Mailstop 70A-3307,\\
Lawrence Berkeley National Laboratory, Berkeley, CA 94720 USA\\
and\\
Institute for Nuclear Theory, University of Washington\\
Seattle, WA 98195-1550}
\date{April 5, 1998}

\maketitle

\begin{abstract}

Since large $p_T$ particles in high-energy hadronic or nuclear collisions
come from jet fragmentation, jet quenching due to parton energy loss in dense
matter will cause the suppression of large $p_T$ hadron spectra in high-energy
heavy-ion collisions. Assuming an effective energy loss $dE/dx$ for the 
high $E_T$ partons, effective jet fragmentation functions are constructed in 
which leading hadrons will be suppressed. Using such effective fragmentation 
functions, high $p_T$ hadron spectra and particle suppression factors 
relative to $pp$ collisions are estimated in central high-energy nuclear 
collisions with a given range of the assumed $dE/dx$. It is found that the 
suppression factors are very sensitive to the value of the effective energy 
loss. Systematic nuclear and flavor dependence of the hadron spectra are also 
studied.
\end{abstract}

\pacs{25.75.-q, 12.38.Mh, 13.87.Ce, 24.85.+p}
]

\section{Introduction}

An ideal quark-gluon plasma (QGP) has often been defined as a system of weakly 
interacting quarks and gluons in both thermal and chemical equilibrium. 
However, recent theoretical investigations based on perturbative 
QCD-inspired model \cite{geiger,XNW97} show that it is increasingly difficult 
for the initially produced partons to evolve into thermal equilibrium, let 
alone chemical equilibrium. Therefore, one might have a generalized 
QGP simply as an interacting and deconfined parton system with a {\it 
large size} and {\it long life-time}. 

One can find many examples of an interacting parton system in collisions 
involving strong interaction. But so far none of them can be considered a QGP
in terms of either the ideal or generalized definition. At a distance much 
smaller than the confinement scale $\Lambda_{\rm QCD}$ and normally in the 
earliest time of the collision, the interaction can be described by 
perturbative QCD (pQCD). Later on, the produced partons will then combine 
with each other via non-perturbative interactions and finally hadronize 
into hadrons.
Therefore, one can consider that there exists an interacting parton system 
during the prehadronization stage in, {\em e.g.}, $e^+e^-$ annihilation 
and deeply inelastic $e^-p$ processes, which is, however, limited only 
to a space-time region characterized by the confinement scale 
$\Lambda_{\rm QCD}$. The characteristic particle spectrum (in $p_T$ and 
rapidity) and the ratios of produced particles are then determined 
by the physics of pQCD and non-perturbative hadronization. 
In ultrarelativistic heavy-ion collisions, one seeks to produce a similar 
interacting parton system but at a much larger scale of the order of a
nucleus size and for a long period of time ({\it e.g.}, a QGP). 
Therefore, one should study those 
experimental observables which are unique to the large size and long 
life-time of an interacting partonic system as signals of a quark-gluon plasma.

Among many proposed signals of a quark-gluon plasma \cite{BM95}, hard 
probes associated with hard processes are especially useful because they 
are produced in the earliest stage of the collision and their abilities 
to probe the dense matter are less complicated by the hadronization physics. 
Merits of hard probes are even more apparent at high energies because 
those processes also dominate
the underlying collision dynamics which will determine the initial
conditions of the produced partonic system \cite{KE97,XNW97}. Study
of them will then enable us to probe the early parton dynamics and the
evolution of the quark-gluon plasma. 

        In general, one can divide the
hard probes into two categories: thermal emission and particle suppression
by the medium. Particle production, like photon/dilepton and charm particles,
from thermal emission can be considered as thermometers of the dense 
medium. Their background comes from the direct production in the initial 
collision processes. On the other hand, suppression of particles produced 
in the initial hard processes, like high-$p_T$ particles from jets 
and $J/\Psi$, can reveal evidences of the parton energy loss in
dense matter and the deconfinement of the partonic system. Thermal
production of these particles is expected to be negligible. Therefore,
in both cases, one needs to know the initial production rate accurately
enough. Another advantage of these hard probes is that the initial production
rate can be calculated via pQCD, especially if we understand the modest
nuclear modification one would expect to happen. 

In this paper, we will discuss high-$p_T$ particles as probes of the dense 
matter since one expect high-$E_T$ partons which produce these high-$p_T$
particles will interact with the dense medium and lose energy.
Medium-induced energy loss of a high-energy parton
traversing a dense QCD medium is interesting because
it depends sensitively on the density of the medium and thus can be 
used as a probe of the dense matter formed in ultrarelativistic heavy-ion
collisions. As recent studies demonstrated \cite{GW1,BDPS,BDMPS}, 
it is very important to take into account the coherent effect in 
the calculation of radiation spectrum induced by multiple scattering
of a fast parton. The so-called Landau-Pomeranchuk-Migdal effect can 
lead to very interesting, and sometimes nonintuitive results
for the energy loss of a fast parton in a QCD medium.
Another feature of the induced energy loss is that it depends
on the parton density of the medium via the final transverse momentum 
broadening that the parton receives during its propagation through the
medium. One can therefore determine the
parton density of the produced dense matter by measuring
the energy loss of a fast parton when it propagates
through the medium.

Unlike in the QED case, where one can measure directly the 
radiative energy loss of a fast electron, one cannot measure 
directly the energy loss of a fast leading parton in QCD.
Since a parton is normally studied via a jet, a cluster
of hadrons in the phase space, an identified jet can
contain particles both from the fragmentation of the
leading parton and from the radiated partons. If we neglect
the $p_T$ broadening effect, the total energy of the
jet should not change even if the leading parton
suffers radiative energy loss. What should be changed
by the energy loss are the particle distributions inside 
the jet or the fragmentation functions and the jet profile.
Therefore, one can only measure parton energy loss indirectly
via the modification of the jet fragmentation functions and
jet profile. For this purpose, it was recently proposed \cite{WHS,WH}
that the jet quenching can be studied by measuring
the $p_T$ distribution of charged hadrons in the opposite
direction of a tagged direct photon. Since a
direct photon in the central rapidity region ($y=0$) is
always accompanied by a jet in the opposite transverse
direction with roughly equal transverse energy, the $p_T$
distribution of charged hadrons in the opposite direction
of the tagged direct photon is directly related to the
jet fragmentation functions with known initial energy.
One can thus directly measure the modification of the
jet fragmentation and then determine the energy loss
suffered by the leading parton with given initial energy.

        Similarly, single-particle spectrum can also be used to 
study the effect of parton energy loss as first proposed in Ref.~\cite{WG92}, 
since the suppression of large $E_T$ jets naturally leads to
the suppression of large $p_T$ particles. However, since the 
single-particle spectrum is a convolution of the jet cross
section and jet fragmentation function, the suppression of
produced particles with a given $p_T$ results from jet
quenching with a range of initial transverse energies.
Therefore, one cannot measure the modification of the jet
fragmentation function or the energy loss of a jet with known initial
transverse energy from the single-particle $p_T$ spectrum as precisely 
as in the case of tagged direct photons. One clear advantage
of single inclusive particle spectrum is the large production rate of 
moderately high $p_T$ particles, while the production rate of large 
$p_T$ direct photons is relatively much smaller at the designed luminosity 
of the Relativistic Heavy-Ion Collider (RHIC) \cite{WH}. Therefore, 
with much less experimental effort, 
one can still study qualitatively the effect of jet quenching and 
extract the average value of the parton energy loss from single 
particle spectra at high $p_T$.

In this paper, we will conduct a systematic study of the effects of parton
energy loss on single-particle transverse momentum spectra in central
$A+A$ collisions in the framework of modified effective jet fragmentation
functions.  We study within this framework the dependence of the spectra
on the effective parton energy loss. We will discuss the energy or $p_T$
and $A$ dependence of the energy loss and jet quenching. Finally, flavor
dependence of the spectra will be also be discussed.

\section{Modified jet fragmentation functions}

Jet fragmentation functions have been studied extensively in $e^+e^-$,
$ep$ and $p\bar{p}$ collisions \cite{mattig}. These functions describe
the particle distributions in the fractional energy, $z=E_h/E_{jet}$, 
in the direction of a jet. The measured dependence of the fragmentation
functions on the momentum scale is shown to satisfy the QCD evolution
equations very well. We will use the parameterizations of the most
recent analysis \cite{bkk,bkk2} in both $z$ and $Q^2$ for jet fragmentation 
functions $D^0_{h/a}(z,Q^2)$ to describe jet ($a$) fragmentation
into hadrons ($h$) in the  vacuum. 

In principle, one should study the modification of jet fragmentation
functions in a perturbative QCD calculation in which induced 
radiation of a propagating parton in a medium and 
Landau-Pomeranchuk-Migdal interference effect can be dynamically
taken into account. However, for the purpose of our current
study, we can use a phenomenological model to describe the
modification of the jet fragmentation function due to an
effective energy loss $dE/dx$ of the parton. In this model
we assume: (1) A quark-gluon plasma is formed with 
a transverse size of the colliding nuclei, $R_A$. A parton 
with a reduced energy will only hadronize outside the deconfined 
phase and the fragmentation can be described as in $e^+e^-$ 
collisions. (2) The mean-free-path of inelastic scattering
for the parton $a$ inside the QGP is $\lambda_a$ which we will keep a
constant through out this paper. The radiative
energy loss per scattering is $\epsilon_a$. The energy
loss per unit distance is thus $dE_a/dx=\epsilon_a/\lambda_a$.

The probability for a parton to scatter $n$ times within
a distance $\Delta L$ is given by a Poisson distribution,
\begin{equation}
  P_a(n,\Delta L)=\frac{(\Delta L/\lambda_a)^n}{n!} 
  e ^{-\Delta L/\lambda_a} \; .
\end{equation}
We also assume that the mean-free-path of a gluon is half
that of a quark, and the energy loss $dE/dx$ is twice that
of a quark. (3) The emitted gluons, 
each carrying energy $\epsilon_a$ on the average,
will also hadronize according to the fragmentation function 
with the minimum scale $Q_0^2= 2.0 $ GeV$^2$. We will also 
neglect the energy fluctuation given by the radiation spectrum
for the emitted gluons. Since the emitted gluons only produce
hadrons with very small fractional energy, the final modified
fragmentation functions in the moderately large $z$ region
are not very sensitive to the actual radiation spectrum and 
the scale dependence of the fragmentation functions for the 
emitted gluons.

This is definitely a simplified picture. In a more realistic scenario,
one should also consider both the longitudinal and transverse expansion.
Because of the expansion, the actual parton energy loss will change as
it propagates through the evolving system resulting in a different total energy
loss as recently studied in Ref.~\cite{baier98}. Since we are mostly interested
in the overall effects, we can neglect the details of the evolution history 
and concentrate on the modification of high $p_T$ hadron spectra due
to an assumed total energy loss or averaged energy loss $dE/dx$ per
unit distance. It might require much more elaborated study to find out
the effects of the dependence of the energy loss on the dynamical
evolution of the system. It is beyond the scope of this paper.

We will consider the central rapidity region of high-energy
heavy-ion collisions. We assume that a parton with initial 
transverse energy $E_T$ will travel in the transverse direction 
in a cylindrical system. With the above assumptions, the modified 
fragmentation functions for a parton traveling a distance $\Delta L$
can be approximated as,
\begin{eqnarray}
  D_{h/a}(z,Q^2,\Delta L)& =&
  \frac{1}{C^a_N}\sum_{n=0}^NP_a(n,\Delta L)\frac{z^a_n}{z}D^0_{h/a}(z^a_n,Q^2)
  \nonumber \\
  &+&\langle n_a\rangle\frac{z'_a}{z}D^0_{h/g}(z'_a,Q_0^2), 
  \label{eq:frg1}
\end{eqnarray}
where $z^a_n=z/(1-n\epsilon_a/E_T)$, $z'_a=zE_T/\epsilon_a$ and
$C^a_N=\sum_{n=0}^N P_a(n)$. We limit the number of inelastic
scattering to $N=E_T/\epsilon_a$ by energy conservation.
{}For large values of $N$, the average number of scattering
within a distance $\Delta L$ is approximately 
$\langle n_a\rangle \approx \Delta L/\lambda_a$.
The first term corresponds to the fragmentation of the 
leading partons with reduced energy $E_T-n\epsilon_a$
and the second term comes from the emitted gluons each
having energy $\epsilon_a$ on the average. Detailed discussion
of this modified effective fragmentation function and its limitations
can be found in Ref.~\cite{WH}.

\section{Energy loss and single-particle $p_T$ spectrum}

To calculate the $p_T$ distribution of particles from jet
fragmentation in $pp$ and central heavy-ion collision, one 
simply convolutes the fragmentation functions with the jet 
cross sections \cite{owens},

\begin{eqnarray}
  \frac{d\sigma_{hard}^{pp}}{dyd^2p_T}&=&K\sum_{abcdh}
  \int_{x_{amin}}^1 dx_a \int_{x_{bmin}}^1 dx_b 
  f_{a/p}(x_a,Q^2) \nonumber \\
  & & f_{b/p}(x_b,Q^2) \frac{D^0_{h/c}(z_c,Q^2)}{\pi z_c}
  \frac{d\sigma}{d\hat{t}}(ab\rightarrow cd), \label{eq:nch_pp}
\end{eqnarray}
for $pp$ and 
\begin{eqnarray}
  \frac{dN_{hard}^{AA}}{dyd^2p_T}&=&K\int d^2r t_A^2(r)\sum_{abcdh}
  \int_{x_{amin}}^1 dx_a \int_{x_{bmin}}^1 dx_b \nonumber \\
  & & f_{a/A}(x_a,Q^2,r)f_{b/A}(x_b,Q^2,r) \nonumber \\
  & & \frac{D_{h/c}(z_c,Q^2,\Delta L)}{\pi z_c}
  \frac{d\sigma}{d\hat{t}}(ab\rightarrow cd), \label{eq:nch_aa}
\end{eqnarray}
for $AA$ collisions, where $z_c=x_T(e^y/x_a +e^{-y}/x_b)/2$, 
$x_{bmin}=x_ax_Te^{-y}/(2x_a-x_Te^y)$,
$x_{amin}=x_Te^y/(2-x_Te^{-y})$, and $x_T=2p_T/\sqrt{s}$.
The nuclear thickness function is normalized to $\int d^2r t_A(r)=A$.
The $K\approx 2$ factor accounts for higher order 
corrections \cite{xwke}. The parton distributions per nucleon in 
a nucleus (with atomic mass number $A$ and charge number $Z$),
\begin{eqnarray}
f_{a/A}(x,Q^2,r)&=&S_{a/A}(x,r)\left[
        \frac{Z}{A}f_{a/p}(x,Q^2)+\right. \nonumber \\
& &\left. (1-\frac{Z}{A})f_{a/n}(x,Q^2)\right],
\end{eqnarray} 
are assumed to be factorizable into
parton distributions in a nucleon $f_{a/N}(x,Q^2)$ and the
parton shadowing factor $S_{a/A}(x,r)$ which we take the
parameterization used in HIJING model \cite{hijing}.
Neglecting the transverse expansion, the transverse
distance a parton produced at $(r,\phi)$ will travel is 
$\Delta L(r,\phi)=\sqrt{R_A^2-r^2(1-\cos^2\phi)}-r\cos\phi$.

In principle, one should also take into account the intrinsic 
transverse momentum and the transverse momentum broadening due to
initial multiple scattering. These effects (so-called Cronin effects)
are found very important to the final hadron spectra at around SPS energies 
($\sqrt{s}=20 - 50 GeV$) \cite{wang98}. However, at RHIC energy which
we are discussing in this paper, one can neglect them (about 10-30\% 
correction) to a good approximation.

\begin{figure}
\centerline{\psfig{figure=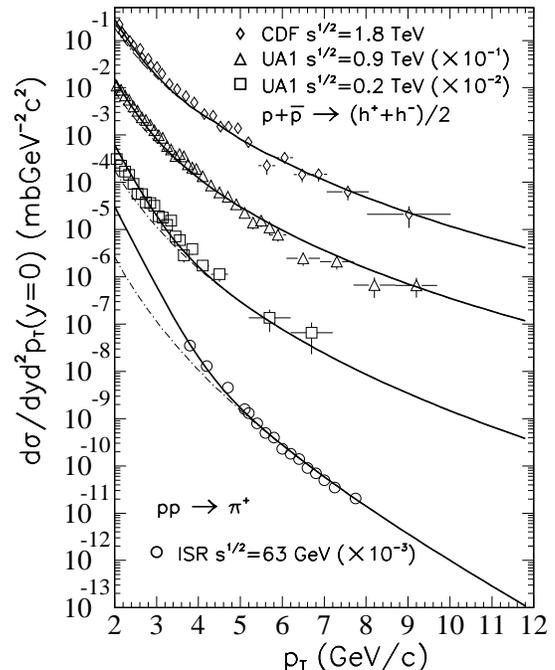,width=2.8in,height=3.5in}}
\caption{ The charged particle $p_T$ spectra in $pp$ and $p\bar{p}$
collisions. The dot-dashed lines are from jet fragmentation only
and solid lines include also soft production parametrized in an
exponential form. The experimental data are from 
Ref.~\protect\cite{alper,ua1pt,cdfpt}.}
\label{figspc1}
\end{figure}

We will use the MRS D$-\prime$ parameterization of the parton
distributions \cite{mrs} in a nucleon. The resultant $p_T$ spectra
of charged hadrons ($\pi^{\pm}, K^{\pm}$) for $pp$ and $p\bar{p}$
collisions are shown in Fig.~\ref{figspc1} together with the
experimental data \cite{alper,ua1pt,cdfpt} 
for $\sqrt{s}=63$, 200, 900 and 1800 GeV.
The calculations (dot-dashed line) from Eq.~(\ref{eq:nch_pp}) 
with the jet fragmentation functions given by Ref.~\cite{bkk,bkk2} 
agree with the  experimental data remarkably well, especially
at large $p_T$. However, the calculations are consistently
below the experimental data at low $p_T$, where we believe
particle production from soft processes, like string fragmentation of
the remanent colliding hadrons, becomes very important.
To account for particle production at smaller $p_T$, we introduce
a soft component to the particle spectra in an exponential form,
\begin{equation}
\frac{dN_{soft}^{pp}}{dyd^2p_T}=Ce^{-p_T/T}, \label{eq:exp}
\end{equation}
with a parameter $T=0.25 $ GeV/$c$. This exponential form is a
reasonable fit to the data of hadron $p_T$ spectra of $p\bar{p}$
collisions at $\sqrt{s}=200$ GeV below $p_T< 2$ GeV/$c$. The fit is
not very good below $p_T=0.5 $ GeV and the parameter $T$ should also
depend on colliding energy $\sqrt{s}$. However, for a rough estimate
of the spectra at low $p_T$ this will be enough and we will keep $T$
a constant.

The normalization in Eq.~(\ref{eq:exp}) is determined from the charged
hadron rapidity density in the central region:
\begin{equation}
C=\frac{1}{2\pi T^2}\left( \frac{dN^{pp}}{dy}
-\frac{dN^{pp}_{hard}}{dy}\right),
\end{equation}
where
\begin{equation}
\frac{dN^{pp}_{hard}}{dy}=\frac{1}{\sigma^{pp}_{in}}\int d^2p_T
         \frac{d\sigma_{hard}^{pp}}{dyd^2p_T}.
\end{equation}
Table~\ref{table} lists the values of the charged hadron rapidity
density and the inelastic cross sections of $pp$ collisions from 
HIJING calculations
which we will use to determine the normalization in Eq.~(\ref{eq:exp})
at different energies.

The total $p_T$ spectrum for charged hadrons in $pp$ collisions
including both soft and hard component is then,
\begin{equation}
\frac{dN^{pp}}{dyd^2p_T}=\frac{dN^{pp}_{soft}}{dyd^2p_T}+
        \frac{1}{\sigma^{pp}_{in}}\frac{d\sigma^{pp}_{hard}}{dyd^2p_T},
\end{equation}
which are shown in Fig.~\ref{figspc1} as solid lines. As one can see
it improves the agreement with data at lower transverse momentum.

\begin{center}
\begin{table}
\begin{tabular}{l l l l l} \hline\hline
 $\sqrt{s}$ (GeV) & 63 & 200 & 900 & 1800 \\ \hline
 $dN^{pp}/dy$     & 1.9 & 2.4 & 3.2 & 4.0 \\ \hline
 $\sigma^{pp}_{in}$ (mb) & 35 & 44 & 50 & 58 \\ \hline\hline
\end{tabular}
\caption{ Charged hadron rapidity density and inelastic cross sections
for $pp$ collisions at different colliding energies from HIJING calculations}
\label{table}
\end{table}
\end{center}

We now also assume that the charged multiplicity from soft particle
production is proportional to the total number of wounded nucleons in
$AA$ collisions which scales like $A$, while the production from hard 
processes is proportional to the number of binary nucleon-nucleon 
collisions which scales like $A^{4/3}$. At
low $p_T$ both type of processes contribute to the particle
spectrum. Therefore the $A$-scaling of the spectrum at low $p_T$
depends on the interplay of soft and hard processes. In HIJING model
\cite{hijing} with a cut-off of $E_{T0}=2$ GeV for jet production the
low-$p_T$ spectra scale like $A^{1.1}$. To take into account of the
uncertainty due to the interplay between soft and hard processes, we
assume hadron spectrum in central $AA$ collisions is,
\begin{equation}
\frac{dN^{AA}}{dyd^2p_T}=A^{\alpha_h}\frac{dN^{pp}_{soft}}{dyd^2p_T}+
        \frac{dN^{AA}_{hard}}{dyd^2p_T},
\end{equation}
where $\alpha_h=1.0 \sim 1.1$.

To calculate $dN^{AA}_{hard}/dyd^2p_T$, we will take into account both
the effect of nuclear shadowing on parton distributions and the
modification of the jet fragmentation functions due to parton energy
loss inside a medium. From Eq.~(\ref{eq:nch_aa}) we see that it will
be proportional to overlap function of central $AA$ collisions
$T_{AA}(0)$. In a hard sphere model for nuclear distribution,
$T_{AA}(0)=9A^2/8\pi R_A^2$ and $R_A=1.2 A^{1/3}$ fm.

\begin{figure}
\centerline{\psfig{figure=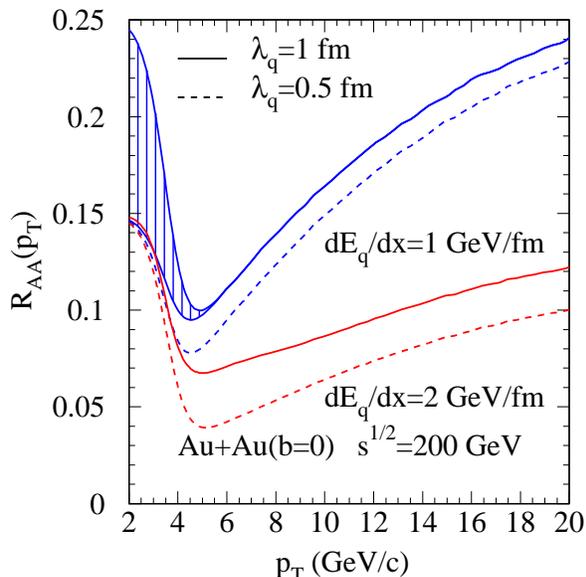,width=3in,height=3in}}
\caption{The suppression factor or ratio of charged particle $p_T$ 
spectrum in central $Au+Au$ over that of $pp$ collisions at 
$\protect\sqrt{s}=200$ GeV , normalized by the total binary nucleon-nucleon
collisions in central $Au+Au$ collisions, with different values of the
energy loss $dE_q/dx$ and the mean-free-path $\lambda_q$ of a quark 
inside the dense medium, the shaded area indicates the uncertainty of
the effective $A$-scaling of low-$p_T$ spectra depending on the
interplay of soft and hard processes}
\label{figspc2}
\end{figure}

We now define an effective suppression factor, or the ratio,
\begin{equation}
  R_{AA}(p_T)=\frac{dN_{AA}/dy/d^2p_T}{\sigma_{in}^{pp}T_{AA}(0)
    dN_{pp}/dy/d^2p_T} \; , \label{eq:ratio}
\end{equation}
between the spectrum in central $AA$ and $pp$ collisions which
is normalized to the effective total number
of binary $NN$ collisions in a central $AA$ collision.
If none of the nuclear effects (shadowing and jet quenching) 
are taken into account, this ratio should  be unity at
large transverse momentum. Shown in Fig.~\ref{figspc2} are the
results for central $Au+Au$ collisions at the RHIC energy
with $dE_q/dx=1$, 2 GeV/fm, and $\lambda_q=1$ (solid), 0.5 fm (dashed),
respectively. As we have
argued before, jet energy loss will result in the suppression
of high $p_T$ particles as compared to $pp$ collisions. Therefore,
the ratio at large $p_T$ in Fig.~\ref{figspc2} is smaller than one 
due to the energy loss suffered by the jet partons. It, however, 
increases with $p_T$ because of the constant energy loss we have
assumed here. At hypothetically large $p_T$ when the total energy loss
is negligible compared to the initial jet energy, the ratio should
approach to one.

Since there is always a coronal region with an average length of
$\lambda_q$ in the system where the produced parton jets will escape
without scattering or energy loss, the suppression factor can never be
infinitely small. For the same reason, the suppression factor also
depends on the parton's mean-free-path, $\lambda_q$. It is thus
difficult to extract information on both $dE_q/dx$ and $\lambda_q$
simultaneously from the measured spectra in a model independent way. 

At small $p_T$, particles from soft processes (or from 
hadronization of QGP) dominate. The ratio $R_{AA}(p_T)$
is then very sensitive to the $A$-scaling behavior of the soft
particle production. Since we assumed an effective scaling,
$A^{\alpha_h}$ with $\alpha_h=1.0 \sim 1.1$, for the low-$p_T$
particle production, the ratio should approach
to $A^{\alpha_h}/\sigma_{pp}T_{AA}(0)=0.149 \sim 0.253$ at 
small $p_T$ for central $Au+Au$ collisions at the RHIC energy, 
as shown in Fig.~\ref{figspc2}.
Therefore, the shaded area in the figure (we only plotted for one case
of energy loss) should be considered
as one of the uncertainties of the ratio at low $p_T$ associated with the
interplay of contributions from soft 
and hard processes. One presumably can determine
this dependence from future RHIC experimental data.

As we have stated earlier, the Cronin effect due to initial multiple parton
scattering will introduce an uncertainty of 10-30\% effect which can be
narrowed down through a systematic study of $p+p$ and $p+A$ collisions.
Since the concept of parton energy loss can only be applied to high $E_T$
jets, it will in principle only affect the spectra at high $p_T$ where 
contribution from soft production is negligible. At smaller values of 
$p_T<3\sim 4$ GeV/$c$ where soft particle production become important,
the connection between parton energy loss and the hadron spectra becomes
unclear. In this region, the modification of the spectra is driven by parton
and hadron thermalization. Since we approximate the spectra in this region
by an effective $A$-scaling of coherent or semi-coherent particle production,
the suppression factor we show in this paper can only be considered as
semi-quantitative. Furthermore, the spectra just above $p_T\sim 4$ GeV/$c$
should also be sensitive to the energy dependence of the energy loss as we
will show in the next section.

\begin{figure}
\centerline{\psfig{figure=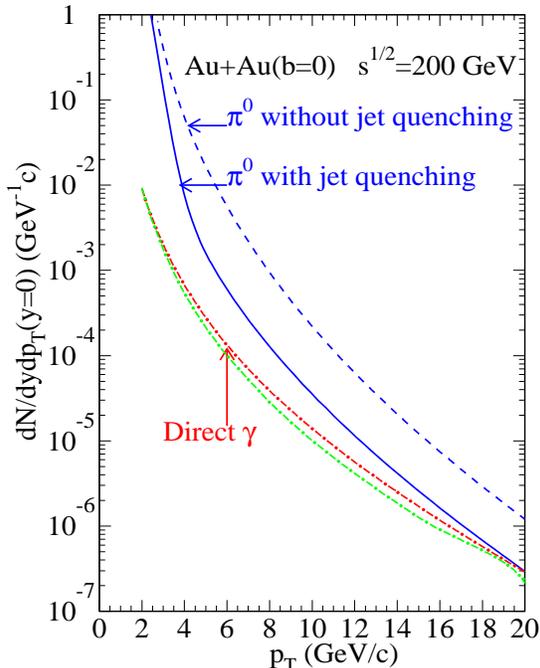,width=2.8in,height=3.5in}}
\caption{The inclusive $p_T$ distribution for $\pi^0$ with (solid) and
without (dashed) parton energy loss as compared to that of direct 
photons (dot-dashed) in central $Au+Au$ collisions at 
$\protect\sqrt{s}=200$ GeV. $dE_q/dx=1$ GeV/fm and mean-free-path 
$\lambda_q=1$ fm are assumed. Contribution from soft particle 
production to the $\pi_0$ spectra is assume to have a $A^{1.1}$ scaling.}
\label{figspc3}
\end{figure}

To further illustrate the effect of the parton energy loss in hadron spectrum
we show in Fig.~\ref{figspc3} the production rates of $\pi^0$ with
(solid line) and without parton energy loss (dashed lines), together
with the spectrum of direct photons (dot-dashed lines) at the RHIC
energy. The upper curve for direct photons is a leading order calculation
multiplied by $K=2$ factor. The lower curve is the result of a next-to-leading
order calculation \cite{cley} which also includes quark bremsstrahlung. 
Here we assumed the low-$p_T$ soft particle spectra scales like $A^{1.1}$.
Since we can neglect any electromagnetic interaction between the produced 
photon and the QCD medium, the photon spectrum will not be affected by the
parton energy loss. On the other hand, jet quenching due to parton
energy loss can significantly reduce $\pi^0$ rate at large $p_T$.
Therefore the change of $\gamma/\pi^0$ ratio at large $p_T$ can also
be an indication of parton energy loss. One can consider the contribution
to direct photon production from bremsstrahlung as quark fragmentation into
a photon, it should in principle also be affected by the quark energy loss
inside the dense medium. Therefore, there is also some uncertainties (
maximum factor of 2 if the $K=2$ factor completely comes from bremsstrahlung
correction) to the estimated photon spectra at lower $p_T$ where
bremsstrahlung is more important.

\section{Energy and $A$ dependence of energy loss}

In recent theoretical studies of parton energy loss \cite{GW1,BDPS,BDMPS}, 
it has been demonstrated that the so-called Landau-Pomeranchuk-Midgal (LPM)
coherent effect can lead to interesting and sometimes nonintuitive
results. Baier {\it et al} have systematically studied these effects in
detail \cite{BDPS,BDMPS}. They found that because of the modification
of the radiation spectrum by the LPM coherence, the energy loss
experienced by a fast parton propagating in an infinite large medium has
a nontrivial energy dependence,
\begin{equation}
        \frac{dE}{dx}\propto
        -N_c\alpha_s\sqrt{E\frac{\mu^2}{\lambda}}
        \ln\frac{E}{\lambda\mu^2} \;\;\;({\rm for} L>L_{cr}),
\end{equation}
where $N_c=3$, $E$ parton's energy, $\mu^2$ the Debye screening mass
for the effective parton scattering, $\lambda$ parton's
mean-free-path in the medium, and $L_{cr}=\sqrt{\lambda E/\mu^2}$.
For a more energetic parton traveling through a medium with finite
length ($L<L_{rc}$), the final energy loss becomes almost independent
of the parton energy and can be related to the total transverse
momentum broadening acquired by the parton through multiple
scattering,
\begin{equation}
        \frac{dE}{dx}=-\frac{N_c\alpha_s}{8} \Delta p_T^2
                =\frac{N_c\alpha_s}{8} \delta p_T^2 \frac{L}{\lambda},
\end{equation}
where $\delta p_T^2$ is the transverse momentum kick per scattering the
parton acquires during the propagation. Therefore, the energy loss per
unit distance, $dE/dx$, is proportional to the total length that the
parton has traveled. Because of the unique coherence effect, the
parton somehow knows the history of its propagation. 

These are just two extreme cases of parton energy and the medium
length. Since it involves two unknown parameters of the medium, it is
difficult to determine which case is more realistic for the system of
dense matter produced in heavy-ion collisions. We will instead study
the phenomenological consequences of these two cases in the final
single inclusive particle spectrum at large $p_T$.

\begin{figure}
\centerline{\psfig{figure=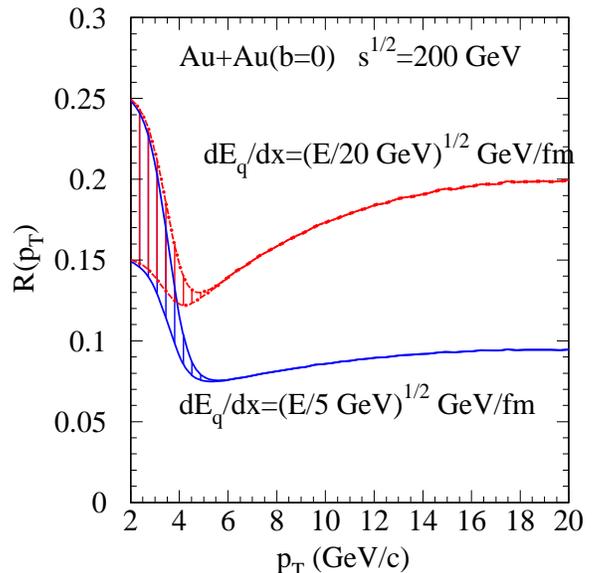,width=3in,height=3in}}
\caption{ The same as Fig.~\ref{figspc2}, except that an
energy-dependent energy loss is assumed. The mean-free-path 
$\lambda_q=1$ fm is used in the calculation.}
\label{figspc4}
\end{figure}

Shown in Fig.~\ref{figspc4} are the calculated suppression factors
with an energy-dependent parton energy loss, 
$dE_q/dx=\sqrt{E/5{\rm GeV}}$ and $dE_q/dx=\sqrt{E/20{\rm GeV}}$
GeV/fm, respectively for central $Au+Au$ collisions at the RHIC
energy. Comparing to Fig.~\ref{figspc2} with a constant energy loss, 
suppression factors are flatter as functions of
$p_T$. This is understandable because the energy loss for larger $E_T$
jet will lose more energy in this scenario thus leading to a stronger
suppression of high $p_T$ particles. As pointed out in
Ref.~\cite{WHS}, the most relevant quantity in the modification  of
the fragmentation functions is the parton energy loss $\Delta E_T$
relative to its original energy $E_T$. For a constant energy loss
$\Delta E_T$, the ratio $\Delta E_T/E_T$ becomes smaller for larger $E_T$,
thus the suppression factor $R_{AA}(p_T)$ will increase with $p_T$. 
If the energy loss $\Delta E_T$ rises with the initial energy
$E_T$, then the increase will be slower. Thus the slope of the ratio
$R_{AA}(p_T)$ can provide us information about the energy dependence
of the energy loss, as one can see from the comparison of
Figs.~\ref{figspc2} and \ref{figspc4}.

\begin{figure}
\centerline{\psfig{figure=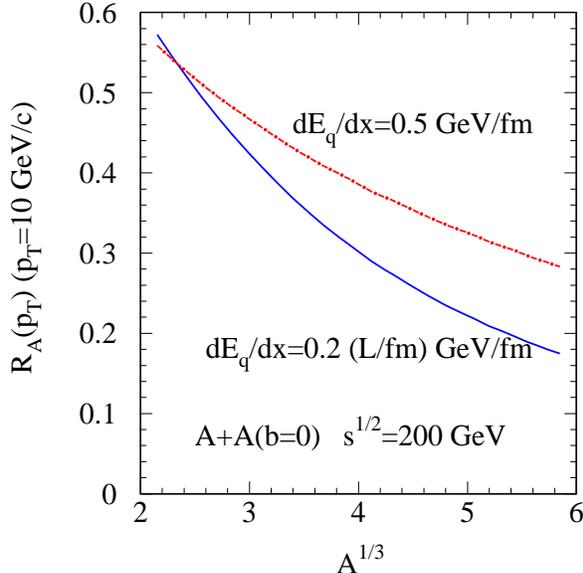,width=3in,height=3in}}
\caption{ The suppression factor for central $A+A$ collisions at
$p_T$=10 GeV/$c$, as a function of the system size, $A^{1/3}$, for a
constant energy loss per unit distance length (dot-dashed line) and 
an energy loss which increases linearly with the length (solid line). 
The mean-free-path $\lambda_q=1$ fm is used.}
\label{figspc5}
\end{figure}

To study the consequences of a parton energy loss $dE/dx$ which
increases with the distance $L$ it travels, one can either vary the
impact-parameter or the atomic mass of the projectile and target so as
to change the size of the dense matter through which the leading
partons have to propagate. Assuming a transverse size of the colliding
nuclei which have a hard sphere nuclear distribution, one can estimate
that the averaged distance a produced jet has to travel through is $\langle
L\rangle_A=1.09 A^{1/3}$, where one has to weight with the probability
of jet production or the overlapping functions of $AA$ collisions.
In Fig.~\ref{figspc5}, we plot the
suppression factor $R_{AA}(p_T)$ at a fixed $p_T$=10 GeV/c for central
$A+A$ collisions at the RHIC energy as a
function of $A^{1/3}$, $A$ the atomic masses of the projectile and
target nuclei. The solid line is for an energy loss, $dE/dx=0.2
(L/{\rm fm})$ GeV/fm, which is proportional to the total length
traveled by the parton, and dot-dashed line is for a constant
$dE/dx=0.5$ GeV/fm. As the size of the system increases, a parton will
lose more energy and thus will lead to increased suppression in both
cases. For collisions of heavy nuclei ( $A^{1/3}>3$), the energy loss
in the first case becomes larger than the second one and thus leads to
more suppression. However, the functional form of the $A$-dependence of
the suppression factor in the two cases do not differ dramatically. It
is therefore difficult to determine whether the energy loss per unit
length is proportional to the total length simply from the 
$A$-dependence of the suppression factor. It must require a model
dependent phenomenological study of the experimental data.

\section{Flavor dependence}

Because of the non-abelian coupling, gluons in QCD always have stronger
interaction than quarks. The gluon density inside nucleons at small
$x$ is larger than quarks; gluon-gluon scattering cross section is
larger than quark-quark; and a gluon jet produces more particles than a
quark jet. For the same reason, a high energy gluon will also lose
more energy than a quark propagating through a dense
medium. Theoretical calculations \cite{GW1,BDPS,BDMPS} all show that
gluons lose twice as much energy as quarks. In this section we will
discuss how to observe such difference in the final hadron spectrum.

By charge and other quantum number conservation, fragmentation 
functions of a gluon jet into particle and anti-particle will be
identical, though it produces more particles than a quark jet and consequently
its fragmentation functions are often softer than a quark's, as has been
measured in the three-jet events of $e^+e^-$ annihilation \cite{gjet}. 
For example, equal number of protons and anti-protons
will be produced in the gluon fragmentation. On the other hand, an
up or down quark is more likely to produce a leading proton than 
anti-proton and vice versa for anti-quarks. Since there will be more
quark (up and down) jets produced than anti-quark in nuclear
collisions, one will find more protons than anti-protons, especially
at large $p_T$ since valence quarks are distributed at relatively
large $x$ (partons' fractional momenta of the nucleon) while gluons
at small $x$. In other words, high $p_T$ protons will have smaller
relative contribution from gluon jets than anti-protons. If gluon jets
lose more energy than quark jets as we have assumed in this paper, one
should then have different suppression factors for proton and
anti-proton. Such flavor dependence should be most evident for heavy
particles like nucleons and lambdas whose fragmentation functions from
a valence quark are significantly harder (i.e., falls off more slowly
at large $z$) and are very different from gluons and sea quarks. For
light mesons like pions, the valence quark fragmentation functions are
softer and are not much different from gluons and sea quarks. One then
will not see much difference between the suppression factors for $\pi^+$
and $\pi^-$ even though gluons and quarks have different energy loss.

\begin{figure}
\centerline{\psfig{figure=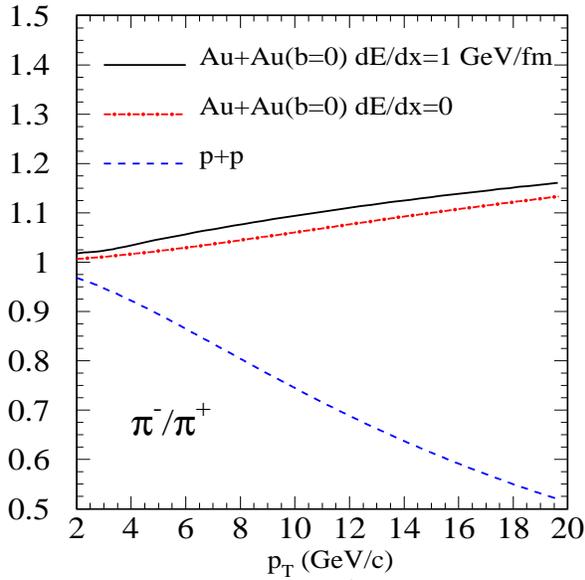,width=3in,height=3in}}
\caption{ The ratio of $\pi^-$ to $\pi^+$ spectra as functions of
$p_T$ in $pp$ (dashed), central $Au+Au$ collisions at $\sqrt{s}=200$
AGeV without energy loss
(dot-dashed) and with energy loss of $dE_q/dx=1$ GeV/fm (The
mean-free-path $\lambda_q=1$ fm).}
\label{figspc6}
\end{figure}

Before we discuss the suppression factors, let us look at the flavor
dependence of the spectra first. Plotted in Fig.~\ref{figspc6} are
$\pi^-/\pi^+$ ratios as functions of $p_T$ in $pp$, central $Au+Au$
collisions with and without energy loss at $\sqrt{s}=200$ AGeV. 
Because gluon-quark scattering dominates in this $p_T$ region at the
RHIC energy and there are twice as
much valence $u$-quarks than $d$-quarks in $pp$ collisions, this ratio
decrease with $p_T$ (dashed line) and should saturate at about 0.5 
(valence $d$ to $u$-quark ratio in a proton) at very high $p_T$ where 
only valence quarks contribute to pion production. At low
$p_T$ where contributions from sea quarks and gluons become more
important the ratio is then close to one. This is a clear prediction
of QCD parton model and has been verified by experiments some years
ago \cite{flvr}. In $Au+Au$ collisions, however, there are slightly more
valence $d$-quarks than $u$-quarks since the nuclei are slightly
neutron rich. As we see in the figure, the $\pi^-/\pi^+$ ratio
(dot-dashed line) then increases with $p_T$ and approaches to a value 
of about $1.14$ which is the valence $d$ to $u$-quark ratio in a $Au$ 
nucleus. The reason why the ratio is different from the limit of $d/u$
ratio is because of finite contributions from sea quarks and gluons. If
gluons lose more energy than quarks, the contribution to high $p_T$
pion production from gluons will be reduced relative to
quarks. Therefore, the $\pi^-/\pi^+$ ratio will be higher than the
case of no difference in energy loss between quarks and gluons (or no
energy loss) or become closer in value to the $d/u$ ratio, as we see 
in the figure (solid line). However, the change due to the parton 
energy loss is very small because the contributions to pion production
from gluons is relatively much smaller than quarks.

\begin{figure}
\centerline{\psfig{figure=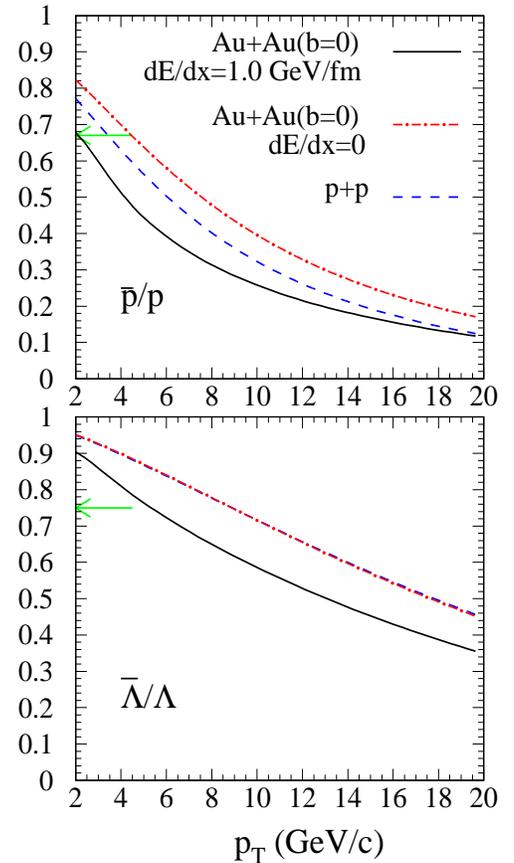,width=2.5in,height=4.5in}}
\caption{ The ratio of $\bar p$ to $p$ (upper panel) and $\bar
\Lambda$ to $\Lambda$ (lower panel) spectra as functions of
$p_T$ in $pp$ (dashed), central $Au+Au$ collisions at $\sqrt{s}=200$
AGeV without energy loss (dot-dashed) and with energy loss of 
$dE_q/dx=1$ GeV/fm (solid) (The mean-free-path $\lambda_q=1$ fm). 
Gluons are assumed to loss as twice much energy as quarks. The arrows 
indicate the ratio at low $p_T<1$ GeV/$c$ from HIJING/BJ estimate (with 
baryon junction model of baryon stopping).}
\label{figspc7}
\end{figure}

        The situation for protons and anti-protons is different. From
parton distributions in a proton we know that gluon to quark density
ratio $f_{g/p}(x,Q^2)/f_{q/p}(x,Q^2)$ decreases with
$x=2E_T/\sqrt{s}$, where $E_T$ is the transverse momentum of the
produced jet.  Consequently the ratio of gluon to quark jet production
cross section always decrease with $E_T$. Since most of anti-protons
come from gluons while protons come from both valence quark and gluon
fragmentation, the ratio of anti-proton to proton production cross
section should also decreases with their $p_T$, as our calculation
shows in Fig.~\ref{figspc7} for $pp$ collisions (dashed line) at
$\sqrt{s}=200$ GeV. At small
$p_T$ gluon and quark jet cross sections become comparable, so the
ratio $\bar{p}/p$ should increase. But it will always be smaller than 1
because there will always be more proton than anti-proton in nucleon
or nuclear collisions due to baryon number conservation (and finite
net baryon production in the central region even from perturbative QCD
calculation). 

\begin{figure}
\centerline{\psfig{figure=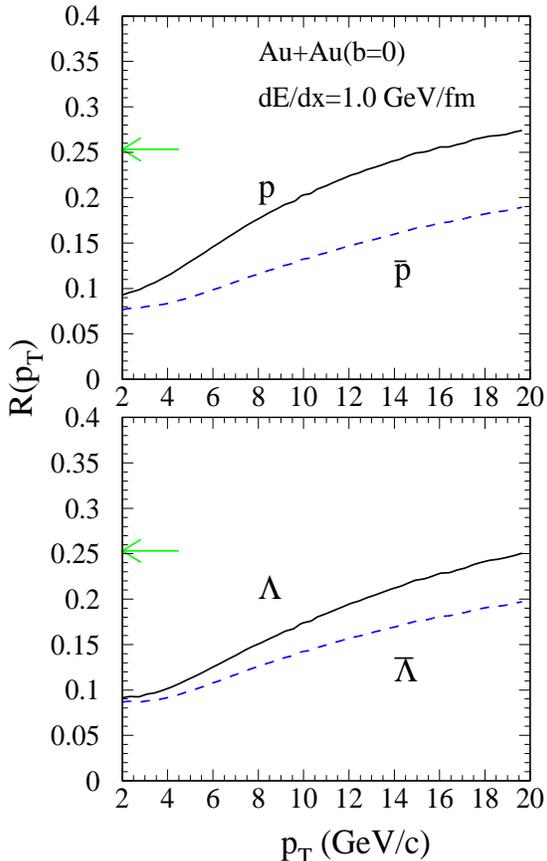,width=2.8in,height=4.5in}}
\caption{ Particle suppression factors for $p$, $\Lambda$ (solid) and 
$\bar p$, $\bar \Lambda$ (dashed) as functions of $p_T$ in central 
$Au+Au$ collisions at $\sqrt{s}=200$ AGeV with energy loss of 
$dE_q/dx=1$ GeV/fm and mean-free-path $\lambda_q=1$ fm. Gluons are 
assumed to loss as twice much energy as quarks. The arrows indicate
the suppression factors at low $p_T<1$ GeV/$c$ if the soft particle
production is assumed to have an $A^{1.1}$ scaling.}
\label{figspc8}
\end{figure}

        The dot-dashed line in Fig.~\ref{figspc7} is the $\bar{p}/p$
ratio in central $Au+Au$ collisions without parton energy loss at the 
RHIC energy. Since $Au$
nuclei are slightly neutron rich, one should have less proton
production per nucleon from valence quark fragmentation
than $pp$ collisions. Since gluon jet production does not change from
$pp$ to $Au+Au$, the ratio $\bar{p}/p$ in $Au+Au$ (without energy
loss) is then a little
larger than in $pp$ collisions. If there is parton energy loss and
gluons lose more energy than quarks, then as we have argued that
$\bar{p}/p$ ratio should become smaller than without energy loss (or
gluons and quarks have the same energy loss), as shown in the figure
as the solid line. The result and argument is the same for
$\bar{\Lambda}/\Lambda$ ratio as also shown in the lower panel of
Fig.~\ref{figspc7}. To further illustrate this point, we plot in
Fig.~\ref{figspc8} the particle suppression factors for proton,
anti-proton, lambda and anti-lambda as functions of $p_T$. Because of
the increased energy loss for gluons over quarks, the suppression
factors for anti-protons and anti-lambdas is then smaller than protons
and lambdas. This could be easily verified if one can identify these
particles at high $p_T$ in experiments.

In the calculation of high $p_T$ baryon spectra in
Figs.~\ref{figspc7} and \ref{figspc8} one has to use parametrized 
fragmentation functions for baryons similarly to these of mesons
\cite{bkk,bkk2}. Though baryon production from jet fragmentation
in $e^+e^-$ and $e^-p$ collisions has been studied\cite{imr},
we could not find any parametrized form including the $Q^2$
evolution. Since Lund model has been proven to reproduce the
experimental data well, here we use the baryon fragmentation functions
parametrized from the Monte Carlo simulation of Lund model
(JETSET) \cite{jetset}. The parameterizations are given in the Appendix.

In the calculation of the particle ratio in Fig.~\ref{figspc6} and 
\ref{figspc7}
and particle suppression factors in Fig.~\ref{figspc8}, we included
only contributions from perturbative hard processes. As we have
discussed before there will also be particle production from
non-perturbative processes. These soft particle production which are
dominant at low $p_T$ are not likely to change the $\pi^-/\pi^+$ ratio
much. However, it might change the $\bar{p}/p$ and
$\bar{\Lambda}/\Lambda$ ratio, as recent heavy-ion experiments show
\cite{sps-p} that there is significantly more baryon stopping than either
pQCD calculation or a simple Lund string model of nuclear collisions.
There are many models of non-perturbative baryon stopping in nuclear
collisions \cite{capella,rqmd}. To take into account of this
non-perturbative baryon stopping, a baryon junction model
\cite{kharzeev} has been implemented into the original HIJING model to
describe the observed baryon stopping at SPS energy \cite{vance}.
This version of HIJING (or HIJING/BJ) model at the 
RHIC energy gives a ratio of
$\bar{p}/p=0.67$ and $\bar{\Lambda}/\Lambda=0.75$. These values should
serve as an estimate of the particle ratio at small $p_T<1$ GeV/$c$,
as indicated by the arrows in Fig.~\ref{figspc7}. This then gives us
an upper bound of the uncertainty for the ratio at around $p_T \sim 2$
GeV/$c$, depending on the interplay between perturbative and
non-perturbative contributions.  Similarly, the suppression factors at
low $p_T$ also depend on the $A$-scaling of low-$p_T$ particle
production. If we assume an $A^{1.1}$ scaling like we did for all
charged particles, then the suppression factors at low $p_T<1$ GeV/$c$
should be as indicated by the arrows in
Fig.~\ref{figspc8}. Since contributions from the stopped baryon scale
like $A$, the low $p_T$ limit for baryons will then be smaller than
anti-baryons. This is an upper bound of the uncertainty one
should bear in mind at intermediate $p_T$. At large $p_T$ these
uncertainties will become very small.

        Similarly as we have discussed in the previous sections, the
particle ratio and suppression factors will all depend on the other
parameters of the energy loss and its energy and $A$ dependence. But
these will not change the qualitative feature of the flavor dependence
of the particle suppression due to different energy loss suffered by
gluons and quarks. Because of finite net baryon density in the central
region, the baryon and anti-baryon absorption in the hadronic phase
will be slightly different which might also give rise to different 
suppression factors for baryons and anti-baryons. 
Detailed study of this effect is
out of the scope of this paper. However, at very large $p_T$, the
physical baryons might only be formed outside the dense region of
hadronic matter. Before then, the color neutral object might have very
small cross section with other hadrons which have already been
formed. Thus the effect of baryon annihilation might be
very small at large $p_T$. Study of the preliminary data for high
$p_T$ particle production in $Pb+Pb$ collisions at the SPS energy 
support this scenario \cite{wang98}.

\section{Conclusions}

A systematic study of the effects of parton energy loss in dense
matter on the high $p_T$ hadron spectra in high-energy heavy-ion
collisions has been carried out in this paper. We found the hadron
spectra at high $p_T$ is quite sensitive to how the large-$E_T$
partons interact with the dense medium and lose their energy before
they fragment into hadrons, leading to the suppression of high-$p_T$
particles. The suppression factor as a function of $p_T$ is also
sensitive to the energy dependence of the parton energy loss. Even
though the nonlinear length dependence of the energy loss as suggested
by a recent theoretical study \cite{BDMPS} leads to stronger suppression,
one cannot unambiguously determine the nonlinearity by varying the
system size. We also studied the flavor dependence of the particle
spectra and the suppression factor and found that it is a good probe of
the energy loss, especially the difference between the energy loss of
a gluon and a quark.

Because of our lack of quantitative understanding of energy loss of
the produced high $E_T$ parton jets inside the dense matter in
heavy-ion collisions, our phenomenological study in this paper can
only be qualitative. But even such a qualitative study is essential to
establish whether there is parton energy loss at all in heavy-ion
collisions and thus whether there is such an initial stage in the
collisions when the produced dense matter is equilibrating. The
analysis we proposed in this paper, which is also somewhat model
dependent, can at least provide information about the average total
energy loss the parton could have suffered during its interaction with
the medium. Anything beyond that will require our knowledge of the
dynamical evolution of the system. Even toward such a modest goal,
there is still one final hurdle to overcome, {\it i.e.}, final state
interactions between leading hadrons of a jet and the soft particles
in the hadronic matter. Such an issue is very important to the
determination whether the high-$p_T$ particle suppression, if any, is
indeed caused by parton energy loss in the initial stage of a dense
partonic matter. Since the formation time of a large $p_T$ particle is
longer than the soft ones, a physical large-$p_T$ particle might be
formed outside the dense region of the hadronic phase. Before then, it
is in a form a color dipole which might have very small interaction
cross section with other hadrons. Therefore, the hadronic phase of the
dense matter might have very small effect on the high-$p_T$ particle
spectra. One could address this issue in heavy-ion collisions at the
SPS energies \cite{wang98}, where one would at least expect that a 
dense hadronic matter has been formed.

\section*{Appendix}

In this Appendix we list the baryon fragmentation functions
parametrized from the results of Lund JETSET Monte Carlo program
\cite{jetset}. We simulate the fragmentation of a $q\bar{q}$ or a
two-gluon system with invariant mass $W=2Q$, and then parameterize the
particle distributions along one direction of the jet axes as functions
of $z=E_h/Q$. We choose the form of the parameterization as \cite{bkk},
\begin{equation}
        D_a^h(z,Q)=\langle n_h(Q)\rangle N
        z^{\alpha}(1-z)^{\beta}(1+z)^{\gamma},
\end{equation}
if the parton $a$ is a gluon or sea quark, and
\begin{eqnarray}
        D_a^h(z,Q)&=&\langle n_h(Q)\rangle [ N_1
        z^{\alpha_1}(1-z)^{\beta_1}(1+z)^{\gamma_1}  \nonumber \\
        &+& N_2 z^{\alpha_2}(1-z)^{\beta_2}],
\end{eqnarray} 
if the parton is a valence quark of the hadron $h$.
The fragmentation functions are normalize as 
$\int_0^1dz D_a^h(z,Q)=\langle n_h(Q)\rangle$.
For a rough approximation which is enough for a qualitative study in
this paper, we neglect the change of the shape of distributions
according to the QCD evolution and attribute the energy dependence to
the average multiplicity $\langle n_h(Q)\rangle$, which are parametrized
as,
\begin{eqnarray}
        \langle n_h(Q)\rangle & =& a+b s+ c s^2, \nonumber \\
        s&=&\ln\frac{\ln(Q^2/\Lambda_{\rm
        QCD}^2)}{\ln(Q_0^2/\Lambda_{\rm QCD})},
\end{eqnarray}
where we choose $Q_0=1$ GeV.
\begin{enumerate}
\item Gluons:
\begin{itemize}
        \item $D_g^n=D_g^{\bar{n}}$
        \begin{eqnarray}
        a&=& 0.061, \;\;\; b= 0.147,  \;\;\; c= 0.155 \nonumber\\
        N&=&3.814, \;\;\; \alpha=-0.187, \;\;\; \beta=3.660, \nonumber\\ 
        \gamma&=&-2.231
        \end{eqnarray}

        \item $D_g^p=D_g^{\bar{p}}$
        \begin{eqnarray}
        a&=& 0.047, \;\;\; b= 0.161, \;\;\; c= 0.133 \nonumber\\
        N&=&3.814  \;\;\; \alpha=-0.187, \;\;\; \beta=3.660,\nonumber\\
        \gamma&=&-2.231
        \end{eqnarray}

        \item $D_g^{\Lambda}=D_g^{\bar{\Lambda}}$
        \begin{eqnarray}
        a&=&0.0215 ,  \;\;\; b=0.0454, \;\;\; c=0.0568 \nonumber\\
        N&=&3.378, \;\;\; \alpha=-0.166, \;\;\; \beta= 4.394, \nonumber\\
        \gamma &=& 0.105
        \end{eqnarray}
\end{itemize}

\item $d$ quarks:
\begin{itemize}
        \item $D_d^n$
        \begin{eqnarray}
        a&=&0.0966, \;\;\; b= 0.0419, \;\;\; c=0.1045 \nonumber\\
        N_2&=&1.671, \;\;\; \alpha_2=0.699, \;\;\; \beta_2=1.311  \nonumber\\
        N_1&=& 0.002, \;\;\; \alpha_1=-2.303, \;\;\; \beta_1=6.461, \nonumber\\
        \gamma_1&=&20.225
        \end{eqnarray}

        \item $D_d^p$
        \begin{eqnarray}
          a&=&0.0392,\;\;\; b=0.0356,\;\;\; c=0.0906 \nonumber\\
        N_2&=&1.377, \;\;\; \alpha_2=-0.252, \;\;\; \beta_2=2.142 \nonumber\\
          N_1&=& 0.005, \;\;\; \alpha_1=-2.246, \;\;\; \beta_1=4.464,  \nonumber\\
        \gamma_1& = &-2.141
        \end{eqnarray}

        \item $D_d^{\Lambda}$
        \begin{eqnarray}
          a&=& 0.0098,\;\;\; b=0.0269, \;\;\; c= 0.0250 \nonumber\\
        N&=&0.230,\;\;\; \alpha=-1.027, \;\;\;  \beta=1.962, \nonumber\\
        \gamma& = &3.037     
        \end{eqnarray}

        \item $D_d^{\bar{n}}$
        \begin{eqnarray}
          a&=& 0.0104, \;\;\; b=0.0867, \;\;\; c= 0.0743  \nonumber\\
        N&=&0.318, \;\;\; \alpha=-0.989,\;\;\;  \beta=4.956, \nonumber\\
        \gamma&=&5.186
        \end{eqnarray}

        \item $D_d^{\bar{p}}$
        \begin{eqnarray}
          a&=& 0.0124, \;\;\; b=0.0676, \;\;\; c= 0.0760 \nonumber\\
        N&=&0.318,\;\;\; \alpha=-0.989,\;\;\;   \beta=4.956, \nonumber\\
         \gamma&=&5.186
        \end{eqnarray}

        \item $D_d^{\bar{\Lambda}}$
        \begin{eqnarray}
          a&=& 0.0033,\;\;\; b=0.0232, \;\;\; c= 0.0265 \nonumber\\
        N&=&0.318, \;\;\; \alpha=-0.989, \;\;\; \beta=4.956,\nonumber\\
        \gamma&=&5.186 
        \end{eqnarray}
\end{itemize}

\item $u$ quarks:

        By isospin symmetry: $D_u^{p(\bar{p})}=D_d^{n(\bar{n})}$, 
        $D_u^{n(\bar{n})}=D_d^{p(\bar{p})}$,
        $D_u^{\Lambda(\bar{\Lambda})}=D_d^{\Lambda(\bar{\Lambda})}$ 

\item $s$ quarks:
\begin{itemize}

        \item $D_s^{\Lambda}$
        \begin{eqnarray}
          a&=&0.0706, \;\;\; b= 0.0546, \;\;\; c=0.0113 \nonumber\\
         N_2&=&11.880, \;\;\; \alpha_2=2.790, \;\;\; \beta_2=1.680 \nonumber\\
        N_1&=& 9.55 \times 10^{-5}, \;\;\; \alpha_1=-3.09, \;\;\; 
        \beta_1=8.344, \nonumber\\
        \gamma_1&=&31.74     
        \end{eqnarray}

        \item $D_s^n$
        \begin{eqnarray}
          a&=& 0.0362, \;\;\; b= 0.0228, \;\;\; c= 0.1087 \nonumber\\
        N&=& 0.254, \;\;\; \alpha=-1.0123, \;\;\; \beta=3.506, \nonumber\\
        \gamma&=&4.385
        \end{eqnarray}

        \item $D_s^p$
        \begin{eqnarray}
          a&=&0.0326, \;\;\; b= 0.0149, \;\;\; c= 0.1060 \nonumber\\
        N&=& 0.421,\;\;\; \alpha=-0.867, \;\;\; \beta=3.985, \nonumber\\
        \gamma &=& 3.577
        \end{eqnarray}

        \item $D_s^{\bar{n}}$
        \begin{eqnarray}
          a&=& 0.0123, \;\;\; b= 0.0631, \;\;\;   c=0.0869 \nonumber\\
        N&=&0.410, \;\;\; \alpha=-0.931, \;\;\; \beta= 5.549,  \nonumber\\
         \gamma&=&4.807     
        \end{eqnarray}

        \item $D_s^{\bar{p}}$
        \begin{eqnarray}
           a&=& 0.0135, \;\;\; b= 0.0456, \;\;\;   c= 0.0935  \nonumber\\
        N&=&0.410 ,\;\;\; \alpha=-0.931, \;\;\; \beta= 5.549,  \nonumber\\
         \gamma&=&4.807      
        \end{eqnarray}

        \item $D_s^{\bar{\Lambda}}$
        \begin{eqnarray}
           a&=& 0.00197, \;\;\; b= 0.0331, \;\;\;  c= 0.0174 \nonumber\\
        N&=& 0.238, \;\;\; \alpha= -1.060, \;\;\; \beta= 7.141, \nonumber\\
         \gamma&=&9.106
        \end{eqnarray}
\end{itemize}

\item Anti-quarks:
        By symmetry of charge conjugate:
        $D_{\bar{q}}^{\bar{B}}=D_q^B$
        $D_{\bar{q}}^B=D_q^{\bar{B}}$
\end{enumerate}

\section*{Acknowledgements}
The author would like to thank S. Vance for providing the estimate of
$\bar{p}/p$ and $\bar{\Lambda}/\Lambda$ ratios from HIJING/BJ calculation.
This work was supported by the Director, Office of Energy
Research, Office of High Energy and Nuclear Physics, Divisions of 
Nuclear Physics, of the U.S. Department of Energy under Contract No.\
DE-AC03-76SF00098 and DE-FG03-93ER40792. The author wishes to thank the
Institute for Nuclear Theory for kind hospitality during his stay when
this work was written.

\end{document}